# Acceleration of collimated 45 MeV protons by collisionless shocks driven in low-density, large-scale gradient plasmas by a $10^{20}$ W/cm$^2$, 1 µm laser

P. Antici[2,3], E. Boella[4], S.N. Chen[1,7], D.S. Andrews[8], M. Barberio[2,3], J. Böker[5], F. Cardelli[3], J.L. Feugeas[6], M. Glesser[1,2], P. Nicolaï[6], L. Romagnani[1], M. Scisciò[2,3], M. Starodubtsev[7], O. Willi[5], J.C. Kieffer[2], V. Tikhonchuk[6], H. Pépin[2], L. O. Silva[4], E. d'Humières[6], J. Fuchs[1,7,*]

[1]LULI - CNRS, Ecole Polytechnique, CEA : Université Paris-Saclay ; UPMC Univ Paris 06 : Sorbonne Universités - F-91128 Palaiseau cedex, France
[2]INRS-EMT, 1650, boulevard Lionel-Boulet, J3X 1S2 Varennes (Québec), Canada
[3]INFN and Dipartimento SBAI, Università di Roma "La Sapienza", 00185 Rome, Italy
[4]GoLP/Instituto de Plasmas e Fusão Nuclear, Instituto Superior Técnico, Universidade de Lisboa, 1049-001 Lisboa, Portugal
[5]Institut für Laser-und Plasmaphysik, Heinrich-Heine-Universität, Düsseldorf, Germany
[6]CELIA, Université de Bordeaux, Talence, France
[7]Institute of Applied Physics, 46 Ulyanov Street, 603950 Nizhny Novgorod, Russia
[8]Lawrence Livermore National Laboratory, Livermore, CA 94551, USA

**Abstract:**

A new type of proton acceleration stemming from large-scale gradients, low-density targets, irradiated by an intense near-infrared laser is observed. The produced protons are characterized by high-energies (with a broad spectrum), are emitted in a very directional manner, and the process is associated to relaxed laser (no need for high-contrast) and target (no need for ultra-thin or expensive targets) constraints. As such, this process appears quite effective compared to the standard and commonly used Target Normal Sheath Acceleration technique (TNSA), or more exploratory mechanisms like Radiation Pressure Acceleration (RPA). The data are underpinned by 3D numerical simulations which suggest that in these conditions Low Density Collisionless Shock Acceleration (LDCSA) is at play, which combines an initial Collisionless Shock Acceleration (CSA) to a boost procured by a TNSA-like sheath field in the downward density ramp of the target, which leads to an overall broad spectrum. Experiments performed at $10^{20}$ W/cm$^2$ laser intensity show that LDCSA can accelerate, from ~1% critical density, mm-scale targets, up to 5x10$^9$ protons/MeV/sr/J with energies up to 45(±5) MeV in a collimated (~6° half-angle) manner.

* :julien.fuchs@polytechnique.fr

Laser-driven proton acceleration [1,2] is a field of intense research due to the numerous fundamental [3,4] and applicative [5,6,7,8] prospects these beams offer. In comparison with conventional accelerators, laser-driven sources offer significant improvements including a higher laminarity at the source; greater flexibility of generated ions; a high number of particles per shot; compact acceleration. The best laser-driven proton beams are produced nowadays with near-infrared (0.8-1 µm wavelength), multi-hundred-TW short-pulse (30 fs–1 ps pulse duration) laser systems generating on-target intensities of ~ $10^{19}$-$10^{21}$ W/cm². These lasers work in standard prepulse laser contrast conditions, i.e., have a ratio of $10^6$ –$10^7$ between the main pulse and the preceding amplified spontaneous emission (ASE), although some facilities have improved performances, i.e. ratio of $10^{10}$ [9]. In these conditions, multi-MeV, divergent (~20° half-angle [10]) beams of protons can be routinely generated by irradiating thin micrometric solid foils (see Refs [1,2] and references therein, and see Refs [11,12,13,14,15] for more recent results). This is commonly achieved using the target normal sheath acceleration (TNSA) mechanism [16] that occurs at the target rear surface [1,2].

In TNSA, high-energy electrons (MeV), generated at the front surface of the target by the laser, cross the target and create at the target rear surface an intense sheath [17] with a ~TV/m electrostatic field that accelerates protons and ions from hydro-vapor contaminants. TNSA is at its most efficient when the target rear surface is unheated; otherwise plasma expansion partially shields the surface electrostatic field [18, 19]. In a variant, TNSA can also take place in a partially expanded target having near-critical density [20, 21]. As the laser can propagate fully through such expanded target, fast electron currents generated near the target rear surface form a long-living quasistatic magnetic field there. This field generates an inductive electric field at the rear plasma-vacuum interface that complements TNSA in providing ion acceleration [22,23] in this so-called Magnetic Vortex Acceleration mechanism (MVA).

Departing from TNSA, several novel acceleration schemes have been investigated in order to increase the kinetic energy of the ions. The first of these novel schemes, Radiation-Pressure Acceleration (RPA), has been studied extensively theoretically [24, 25] as well as demonstrated experimentally [14,26,27,28]. A second acceleration scheme that has been emerging is the Burn Out Afterburner (BOA) in which an ultra-intense laser can relativistically penetrate through an ultra-thin target during the laser lifetime while the target is being exploded and in which an instability coupling the electrons accelerated by the transmitted laser and the target ions sets up, which accelerates the latter [29,30]. A third mechanism is the Collisionless Shock Acceleration (CSA) mechanism, initially proposed by Silva et al. [31]. There, protons are accelerated by being pushed from the potential barrier located at a shock front generated by the impinging laser. Such collisionless shock wave is generated following the injection in the target, beyond the critical density interface at which the laser is stopped, of laser-accelerated fast electrons. Due to their high energy, the collisional dissipation onto these electrons is negligible [32], however collisionless (i.e. mediated by instabilities and plasma waves) processes can provide enough energy dissipation [33], such that a density steepening can form as the fast electrons overcome in their propagation the target medium.

For CSA to produce maximum ion energies comparable or above those produced by TNSA, ultra-intense laser pulses are required (I ≥ $3\times10^{20}$ W/cm$^2$) [34, 35]. This mechanism, disentangled from TNSA, has been experimentally demonstrated on a 10 µm wavelength $CO_2$ laser using as target a near-critical $H_2$ gas target [36, 37]. This yielded collimated protons with an energy up to ~20 MeV.

Recently, another acceleration scheme, namely the low-density collisionless shock acceleration mechanism (LDCSA) [38], has been highlighted in simulations. Contrary to the "pure" CSA mechanism revealed in the refs. 30 and 31, which requires the production of a shock in an overcritical density medium and is difficult to achieve homogeneously for a 1 µm wavelength laser, LDCSA combines shock front acceleration with the volumetric low density acceleration variant of TNSA. We stress that in LDCSA the shock is not generated by the laser, but by the protons propagating downward the density gradient. In details, the ions are first accelerated in the same manner as for the inductive variant of TNSA. Due to the smooth density gradient at the rear side, the inductive electric field monotonously decreases with the distance from the high-density zone. Ions in the low-density region therefore experience an electric field lower than ions from the higher density region. As a result, ions from the low-density region can be overtaken by the ions coming from the higher density region leading to the formation of an electrostatic shock front, a peak of ion density propagating inside the decreasing (low) density ramp (as illustrated in the simulations detailed below). The ions located ahead of the shock structure can be reflected by it, and accelerated at velocities up to twice the shock velocity. We note that, as the inductively-accelerated fast ions coming from the high density region have various energies, the shock is broad in the velocity spectrum, and hence the CSA accelerated ions have an overall broad energy spectrum (as observed here).

This mechanism has never been demonstrated experimentally, but simulations have suggested that exploding sub-micrometric foils could offer appropriate plasma conditions for generating high-energy protons through this mechanism [39]. We stress however that such conditions are different from the ones favorable to BOA acceleration: in ultra-thin foils the transparency regime of the laser (BOA) plays a critical role in the acceleration mechanism while in LDCSA the laser is fully absorbed due to the very large scale density gradients involved.

In recent experiments using an intense picosecond laser pulse (I~$5\times10^{18}$ W/cm$^2$), we observed that protons, accelerated using targets that were only partially decompressed, exhibit similar maximum proton energy and number compared to proton beams that are produced, in similar laser conditions, from solid targets using TNSA. However, in the then-accessible laser and plasma conditions, no clear demonstration of LDCSA could be obtained [40].

In this paper we present evidence that in fully exploded, low-density ($10^{19}$ cm$^{-3}$) plasmas irradiated by a high-intensity laser pulse ($10^{20}$ W/cm$^2$), protons are observed to be accelerated to very high energies (the high-energy cutoff is observed at 45±5 MeV repeatedly over three shots performed in the same conditions) and in a highly directional manner (6° FWHM), i.e. to similar energies and in a much narrower beam than that produced by TNSA from solid foils (where we recorded at best 36±2 MeV) using the same laser pulse and laser intensity. Particle-In-Cell simulations performed in the conditions of the experiment suggest that the LDCSA mechanism is at play and effective in such large-scale gradient, low-density plasmas. We also show that the number of accelerated particles produced in these conditions ($5\times10^9$ protons/MeV/sr/J) is comparable, and even higher for high-energy protons, to those obtained in the TNSA regime. This regime therefore provides an interesting alternative to TNSA, one that does not require strong constraints on the laser temporal contrast (like RPA or BOA), that can exploit under-critical plasmas from inexpensive exploded targets, and that can produce a very directional proton beam.

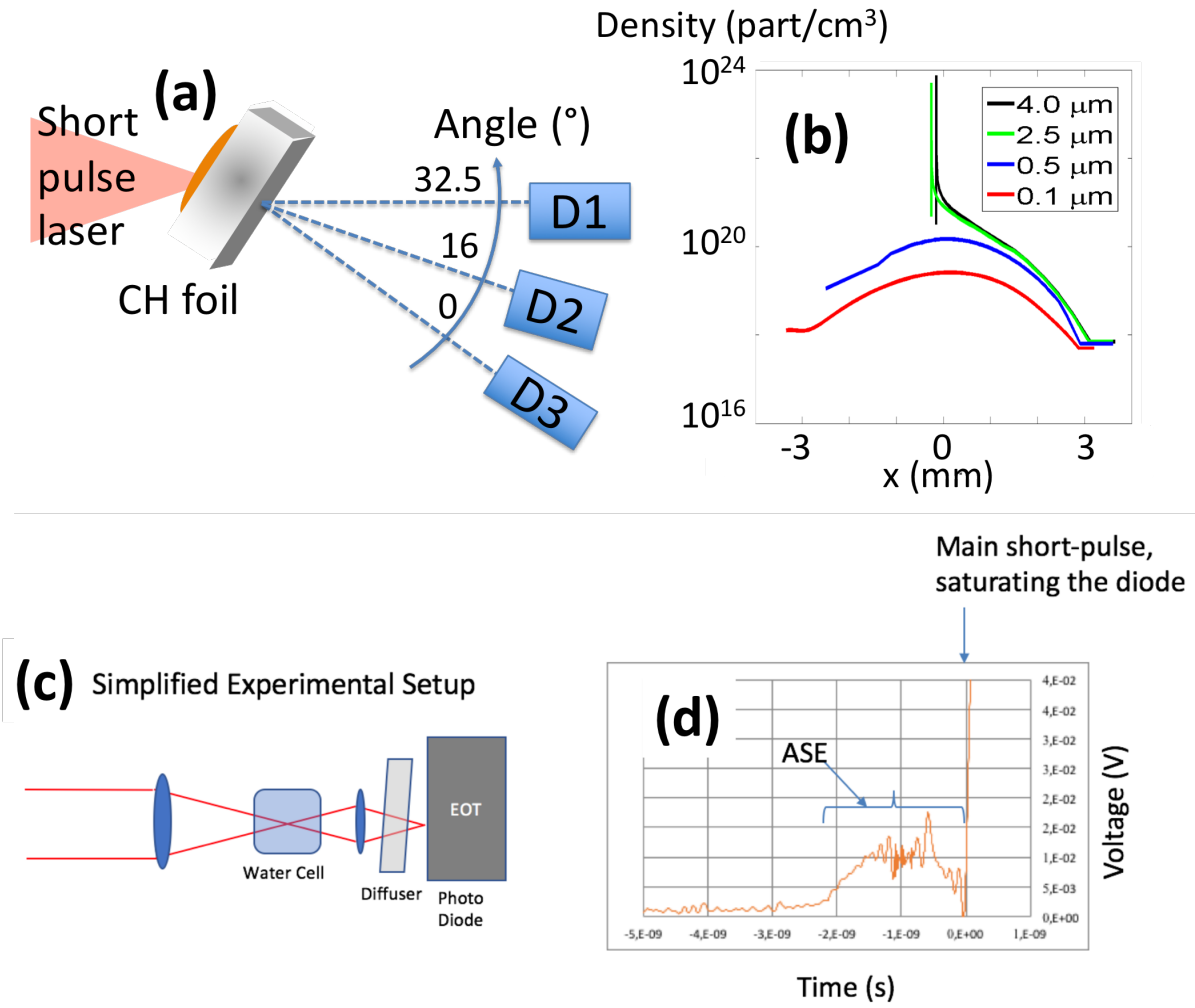

*Figure 1: (a) Setup of the experiment. Targets are either Au, to generate protons in standard conditions using the electrostatic TNSA mechanism, or thin Mylar foils, the thinnest being exploded to low densities by the ASE of the short-pulse. Three spectrometers are used to record the emitted proton spectra with high spectral resolution. A sample of the spectra is shown in Fig.2.b (for the comparative simulations see Fig.3). The spectrometers are located at various angles, as indicated, D3 being placed along the target normal, i.e. at 0°. Complementarily, films were also used on some shots to record the whole angular profile of the protons. A sample of these films is shown in Fig.4. (b) Electron density profiles, as modeled by the 1D Esther hydrodynamic-radiative simulation code [47] resulting from the irradiation of the Mylar targets by the ASE of the short-pulse laser, and just prior to the main short-pulse irradiation. (c) Setup of the ASE (prepulse) measurement of the main beam after*

*compression: a leakage of that beam through a mirror is sent, through a window of the compressor chamber, to a water cell before being collected by a fast photodiode. The water cell serves as absorbing the main compressed pulse part of the signal, preventing the diode to be damaged. On the contrary, the ns-duration, ASE part of the signal can be measured, being not intense enough to be absorbed in the water cell. (d) Oscilloscope trace of the signal collected of one shot of the experiment. Before the saturation induced by the short-pulse part of the signal, one sees the trace of the ASE, having a short ramp (~0.5 ns) preceding a ~1.5 ns flat plateau.*

The experiment was performed using the TITAN laser facility located at the Lawrence Livermore National Laboratory (LLNL), Livermore (USA). Its short-pulse (SP) laser has an energy up to 180 J, a pulse duration of $\tau_{SP}$=700 fs, a wavelength $\lambda_{SP}$=1.054 µm, and a 6-8 µm Full Width Half Maximum (FWHM) focal spot containing ~50% of the laser energy, leading to a peak intensity I~1-2×$10^{20}$ W/cm². The experimental set-up is shown in Fig.1.a. The targets are thin Mylar (PolyEthylene Theraphtalate or PET) foils of different thicknesses ranging from 0.1 to 4 µm; they are irradiated at an angle of 32.5° by the SP. As a basis of comparison, we also irradiated 4, 6, 10, 15 and 25 µm thick solid Au targets in standard TNSA conditions to verify which proton beam parameters could be produced in optimum TNSA conditions.

The thin (0.1 and 0.5 µm thick) Mylar targets are exploded by the ASE of the SP (see Fig.1a). This is due to the fact that the ASE we recorded during the experiment is significant; note that using lasers having a better temporal contrast (improved e.g. by three orders of magnitude), similar targets can be used while keeping a solid density at the time of the main interaction [9], while such targets are fully exploded here.

We also stress that, on a practical point of view, these targets are not only easy, but also inexpensive to manufacture by spin-coating a polymer liquid solution on a disk and then removed by flotation the formed thin Mylar sheet. This is quite different than the targets used so far for RPA studies, which are just a few nm thick and are usually manufactured using micro-lithography techniques.

The ASE that explodes these thin targets has a duration of 2 ns and was measured, during the experimental run, to contain ~10 mJ in energy (at the target chamber center (TCC), i.e. at the location of the SP focus), as measured with fast diodes and a water-switch cell (see Fig.1.c). The calibration of the measurement was made by sending a low-energy, 3 ns duration pulse through the chain and the compressor, and measuring its energy simultaneously at TCC, and on the diode which measures the ASE on every shot (illustrated on Fig.1.c). Overall, the ASE is characterized by an intensity of $10^{13}$ W/cm², a 500 ps rise time, followed by a 1.5 ns plateau, which precedes the peak at the nominal SP (see Fig.1.d). Note that these values are similar to the ones measured in other runs at the same laser facility by other groups [41,42].

The protons accelerated from the target rear are analyzed with three diagnostics, as shown in Fig.1.a, D1 and D3 being Thomson Parabolas [43] and D2 a magnetic spectrometer (not equipped with electric field plates as D1 and D3). The diagnostic D3 is aligned with the target normal, i.e. at the 0° angle, as shown in Fig.1.a. Proton spectra were readout in an absolute manner [44] using TR-type ImagePlates (BAS-TR 2025, Fuji Photo Film Co. Ltd) that were analyzed using a FUJIFILM FLA-7000 reader. The conversion from the units of the Image Plate (PSL) to protons was made using the absolute calibration given in Ref. [44] in order to obtain the spectra shown in Fig.2.b. Since D2 did not have the capability to separate various ion species, we verified that the dominant signal, and in particular the high energies recorded at the cutoff were related to protons by using filters. This was done by adding a 2 mm thick Cu filter right on the ImagePlate. Such filter has, according to SRIM [45,46], a cutoff of: 34 MeV for protons, 750 MeV for C (62 MeV/nuc), and 1.05 GeV for O (65 MeV/nuc). Since the signal near the cutoff emerging from the filter was similar to the one obtained without the filter, we concluded that it was due to protons, otherwise, it would imply even higher energy ions. Additional measurements confirming the proton spectra (see Fig.2.b) and energies (see Fig.2.a) were obtained using RadioChromic Films (RCFs) (MD-55 type) [10], located between 30 and 35 mm (depending on the shots) from the target that allowed measuring the spatial distribution of the beam (shown below).

The expected density profiles of the exploding targets at the end of the ASE irradiation are shown in Fig.1.b. They were obtained using the 1D hydrodynamic ESTHER code [47], benchmarked with the CHIC 2D hydrodynamic code [48]. Here we rely on hydrodynamic simulations, using the well characterized ns-duration, low-intensity ASE of the SP, to infer the target density profiles at the time of the SP irradiation; such procedure has indeed been validated quantitatively many times over the years, as shown e.g. in Refs [40,49,50,51,52]. We can readily see that for thicknesses above 2 µm, the target rear density profiles are steep, which is not favorable for LDCSA to take place [38]. For targets that are thinner, the low-density profiles offer a priori better conditions for LDCSA [38]. This is represented in the cartoon on top of Figure 2.a.

Figure 2.a shows the maximum proton energy recorded by irradiating with the SP these different plasma density profiles. It exhibits two maxima: one in the fully exploded foil regime for the target of 0.1 µm (where the high-energy cutoff is observed at 45±5 MeV repeatedly over three shots performed in the same conditions), and the other for the 2.5 µm thick target (where the high-energy cutoff is observed at 42±5 MeV, also over three shots performed in the same conditions) which is still at solid-density. Note that, as shown in the spectra shown in Fig.2.b, the uncertainty on the cut-off proton energy, at these energies and for a single shot, induced by the dispersion of the spectrometer, is 3 MeV. Comparatively, using 15 µm thick Au targets, where TNSA is dominant [1,2], we obtained a maximum energy of 36±2 MeV (and 30 ±2 MeV at 16° from the target normal) over four shots performed in the same conditions. This was the best performance we obtained when scanning the various Au target thicknesses mentioned above (4, 6, 10, 15 and 25 µm); this optimum point for a 15 µm target and the recorded 36±2 MeV energy is consistent with previous measurements of TNSA-generated proton beams performed also at TITAN by various other groups, see e.g. Refs [42,53,54,55].

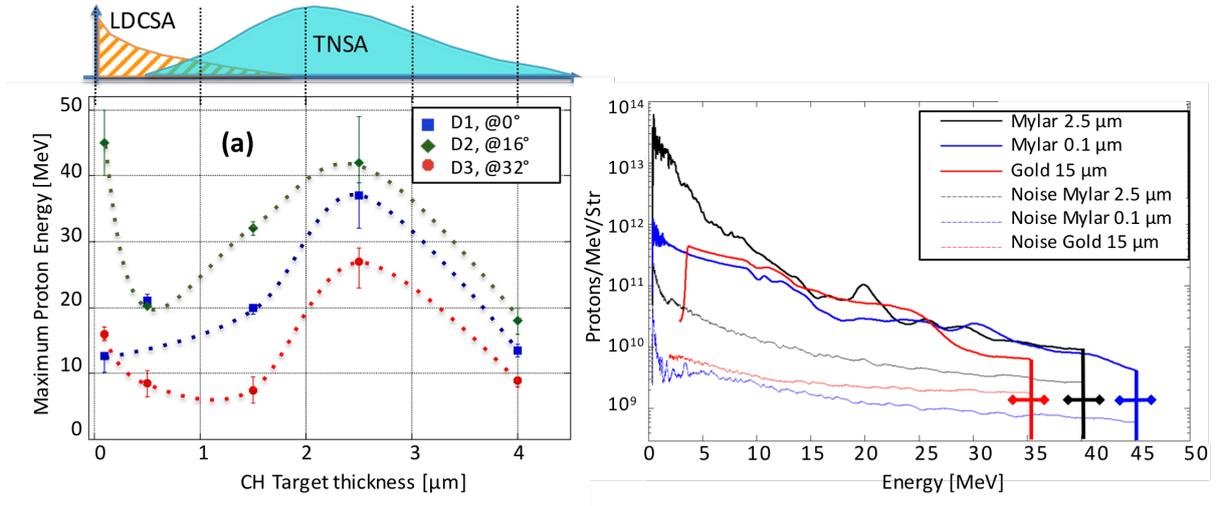

*Figure 2: (a) Maximum proton energy vs initial thickness of the Mylar targets, whose density profile at the time of the main pulse irradiation are shown in Figure 1.b. The lines are guides for the eye. Above is a cartoon showing the predominance of each acceleration regime depending on the plasma density profile (fully exploded targets or solid-density targets). (b) Proton spectra recorded by the D2 magnetic spectrometer (not equipped with electric field plates as D1 and D3) for three target conditions, as indicated. The 0.1 and 2.5 µm are Mylar targets, and correspond to the data shown in (a). The spectrum recorded from a 15 µm thick Au target is also shown as reference of standard proton acceleration in the TNSA regime. The corresponding noise level recorded, on the side of the proton trace, on the ImagePlate, is shown in dashed for each spectrum. The horizontal error bars represent the uncertainty on the cut-off proton energy, at these energies and for a single shot, induced by the dispersion of the spectrometer, which is 3 MeV (represented as ±1.5 MeV around the cut-off).*

The angular distribution of the accelerated protons is also quite different for the two maxima shown in Fig.2.a: the 0.1 µm target produces a much more directional proton beam when compared to the 2.5 µm target. This is consistent with what is observed in the simulations shown in Fig.3 and detailed below. The energy in the 16° direction exceeds 40 MeV while the Thomson Parabolas at other angles (0°, 32°) detect significantly lower proton energies (we verified that the high energy recorded in the magnetic spectrometer D2, not equipped with electrodes as D1 and D3, was indeed due to protons, and not other ions, using filters positioned directly onto the detector and able to block heavier ions than protons, as detailed above). This is confirmed by looking at the 2D beam profile recorded in RCFs as shown in Fig.4: the thick target displays a standard TNSA flat, broad profile [10] while the thin target displays a narrow angular proton beam profile (see also the azimuthally angular proton beam profiles in Fig.4.b). We also see in Fig.2.a that the 2.5 µm target results in significantly less pronounced directional acceleration. In both cases (the 2.5 µm target and the exploded 0.1 and 0.5 µm targets), maximum proton energies are not found for normal incidence but recorded by the spectrometer at 16°, as also confirmed by the RCFs shown in Fig.4 which clearly show that the main proton dose is off-center in these cases, with the proton beam becoming narrower in angle when going to the exploded foils. Evidently, in the RCFs shown in Fig. 4, we observe that the lateral offset of the proton beam center in the exploded foil regime is of 17° with a FWHM of 12°. This is likely due to the impinging ASE from the short-pulse off-axis parabola generating a plasma expansion in the target normal direction. Thus the protons are accelerated within a directional plasma, which tilts the axis of the most energetic protons, as already observed in Ref. [56]. Hence it is expected that the spectrometer D2 will record the highest energy protons corresponding to the center of the beam and that D1 and D3 see only low energy protons because they sample the far-off wings of the beam.

Interestingly, we also observe in Figure 2.b that the proton spectrum recorded for the 2.5 μm Mylar target is quite similar to that of the 15 μm Au target, in contrast to the spectrum recorded for the 0.1 μm Mylar target where the slope is much less steep. Comparing the proton number from all these targets, all spectra at lower energies reach a maximum proton fluence $\geq 10^{12}$ protons/MeV/sr, i.e. $5 \times 10^9$ protons/MeV/sr/J. This is to be compared to $2 \times 10^{11}$ protons/MeV/sr/J obtained by Palmer et al. [36] and $1.5 \times 10^5$ protons/MeV/sr/J obtained by Haberberger et al. [37]. The latter result is much lower in terms of efficiency, but we note that Fiuza et al. [57] suggest that laser filamentation could be responsible for the low charge of the ion bunch in that. The fact that our fluence is lower than the one obtained by Palmer et al. [36] could come from the presence of large scale density gradients in our exploded targets, which were not present in their case as they used a gas jet.

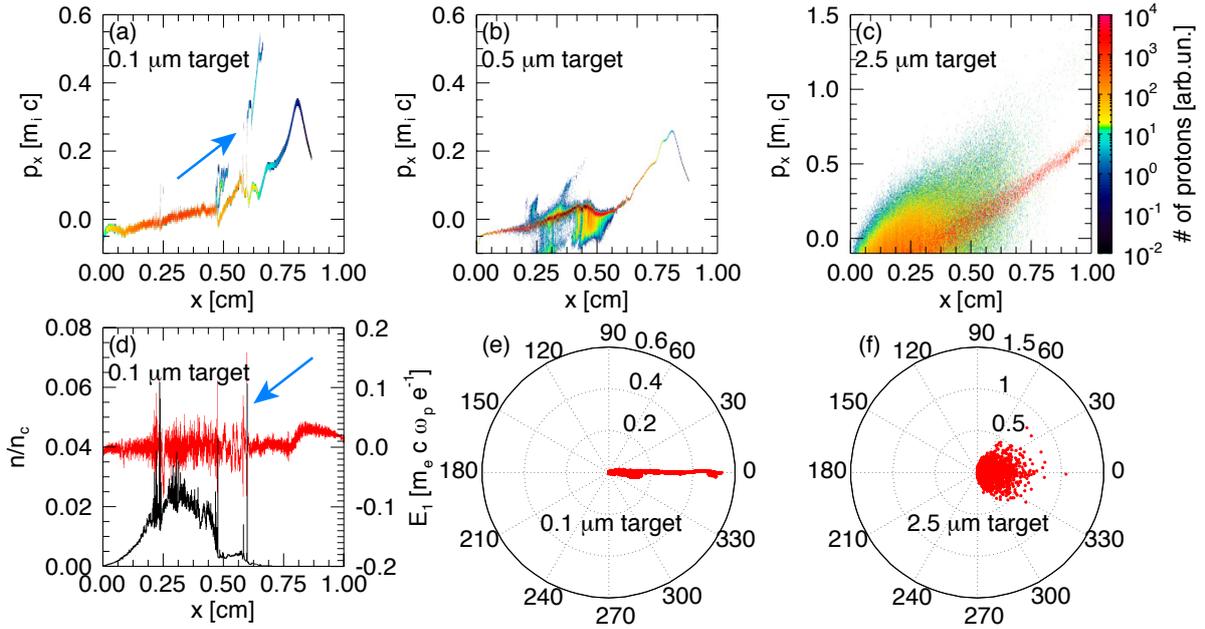

*Figure 3 :* (a-c) 3D PIC simulated proton phase-space (proton momentum vs longitudinal position) at t = 33.7 ps (the laser pulse enters the simulation box on the left side at t = 0) for the 0.1 (a), 0.5 (b) and 2.5 (c) μm targets. The density profiles used as inputs in the simulations are the ones shown in Figure 1.b. (d) Transversely averaged ion density profile (black) and longitudinal electric field (red) corresponding to the case (a). (d) Simulated angular distribution of the accelerated protons in the case of the 0.1 μm target (corresponding to the spectrum shown in (a)). (e) Same as (d) for the 2.5 μm target (corresponding to the spectrum shown in (c)).

To get an insight into the mechanisms responsible for the generation of high-energy protons in the different plasma density profiles used in the experiment, we performed first principles 3D Particle-In-Cell (PIC) simulations with the fully relativistic and massively parallel code OSIRIS 3.0 [58]. In the simulations, pre-formed plasmas composed of Maxwellian electrons with a temperature in agreement with the ponderomotive scaling [59] considering the intensity of the ASE and cold $H^+$ ions ($m_i/m_e$ = 1836), with density profiles described by Fig. 1.b, were introduced. Preliminary simulations performed using a mixture of $H^+$ and $C^{6+}$ ions did not show noticeable differences. The envelope of the SP was modeled with a Gaussian like polynomial profile given by $10\tau^3 - 15\tau^4 + 6\tau^5$, with $\tau = t/t_{FWHM}$ and $t_{FWHM}$ = 700 fs and infinite spot size. We note that the use of a plane wave will obviously lead to an overestimation of the simulated final ion energy and prevents a direct quantitative comparison with experimental data. Nonetheless, as detailed below, since the simulations model the experimental trends correctly with respect to energy and angular distribution, we can infer from the simulations the predominance and interplay between LDCSA and TNSA in the different targets, i.e. exploded or solid-density. The considered laser pulse has a wavelength $\lambda$ = 1 µm and a peak normalized vector potential $a_0 \sim 9$, following the experimental parameters. Since the computations need to be performed over the cm-scale longitudinally, and hence over tens of ps, we had to resort to a small transverse dimension: the 3D simulations used a box size of $10^4 \times 4 \times 4$ µm$^3$. The simulation domain was discretized in cubic cells with $\Delta x$ = 0.08 µm. Periodic boundary conditions were employed for the transverse directions, while absorbing and VPML [60] boundary conditions were utilized for particles and fields respectively along the longitudinal direction. In order to model the plasma dynamics correctly, $10^9$ numerical particles and quadratic interpolation have been used. Particles have been pushed for more than 3.5 x $10^5$ time steps with $\Delta t$ = 0.16 fs = 0.3 $\omega_0^{-1}$, $\omega_0$ being the laser frequency, corresponding to 58 ps, for a total time of about $10^5$ CPUh per simulation. We would like to stress that resorting to 3D simulations has been necessary due to the unsuitability of 2D simulations to capture correctly the interplay between LDCSA and TNSA. Preliminary 2D simulations performed under the same conditions showed indeed that TNSA was always the dominant acceleration mechanism, preventing shock formation and ion reflection.

Figure 3.a-c shows the phase spaces diagrams for three different target profiles representing the two high-energy proton peaks observed in the experiment (i.e. for the 0.1 and 2.5 µm targets) and an intermediate case (0.5 µm) that resulted in lower protons energies. In particular, in Fig.3.a, we can identify, for the 0.1 µm target, a peak at high-energy in the phase space that is generated at early time during the acceleration process and travels towards the back target surface. This peak in the longitudinal momentum ($p_x$) corresponds to a shock that propagates within the target and accelerates the ions pushing them from the front surface to the back surface. This strong shock (located by an arrow) is also visible in Fig.3.d, which shows the ion density at the same time. We note that in our simulations, we did not observe any magnetic field structures. The self consistent magnetic field in all the directions is completely negligible. Hence, we can rule out that the MVA mechanism is at play here.

Other peaks are observable in Fig.3.d: they are secondary non-linear waves that develop during the process and present corresponding non-linear structures in the phase-space. Note that there is also a developed TNSA peak located towards the rear of the target; it does not display the characteristic linear chirp of TNSA-accelerated protons. Unlike the case of standard TNSA, here the longitudinal electric field decreases towards the end of the target, therefore inner ions are accelerated more than the outer ones and overtake them, creating the hump-like feature. This behaviour is similar to what is described in Ref. [61], where such features were observed due to charge separation effects. The fact that TNSA adds to LDCSA acceleration for lower-energy protons results in the overall generation of a continuous spectrum. Note also that the high-energy protons are accelerated in the simulation in a narrow angular range (see Fig.3.e), which is well-consistent with the experimental measurements from this 0.1 μm target, as the spectrometers D1 and D3 record much lower proton energies than the magnetic spectrometer D2.

The same weak TNSA develops in the exploding 0.5 μm target, but the strong shock structure is here noticeably absent (see Fig.3.b). This is well consistent with the experimental results in which the protons are of lower energy in the case of the 0.5 μm target. For the thicker (2.5 μm) target (Fig.3.c) we see a phase space, typical of the TNSA regime, in which the longitudinal momentum increases when going towards the target rear surface since the acceleration occurs mostly there. Note also that the protons accelerated in the 2.5 μm thick target are emitted in a wide angular region (see Fig.3.f), which is again well-consistent with the experimental measurements from the same thick target, as D1 and D2 record very similar spectra in this case. For even thicker targets, i.e. for thicknesses above 4 μm, the target is not sufficiently decompressed and therefore part of the laser-energy is reflected and not absorbed by the target and transferred into the acceleration process. In summary, the proton cut-off energy versus target thickness presents a maximum for very thin targets, where LDCSA is the dominant acceleration mechanism. It then decreases to a minimum when LDCSA strongly competes with TNSA. Finally, when TNSA becomes the dominant mechanism of acceleration, the proton energy increases again till the second maximum. The angular distribution is moreover well reproduced compared to the experimental results. All this shows that the acceleration processes highlighted by the PIC simulations confirm the intuitive picture depicted above Fig.2.a of the transition of a predominance between LDCSA to TNSA as the plasma density profile evolves from a low-density, expanded one, to one having a solid-density with a sharp rear surface gradient.

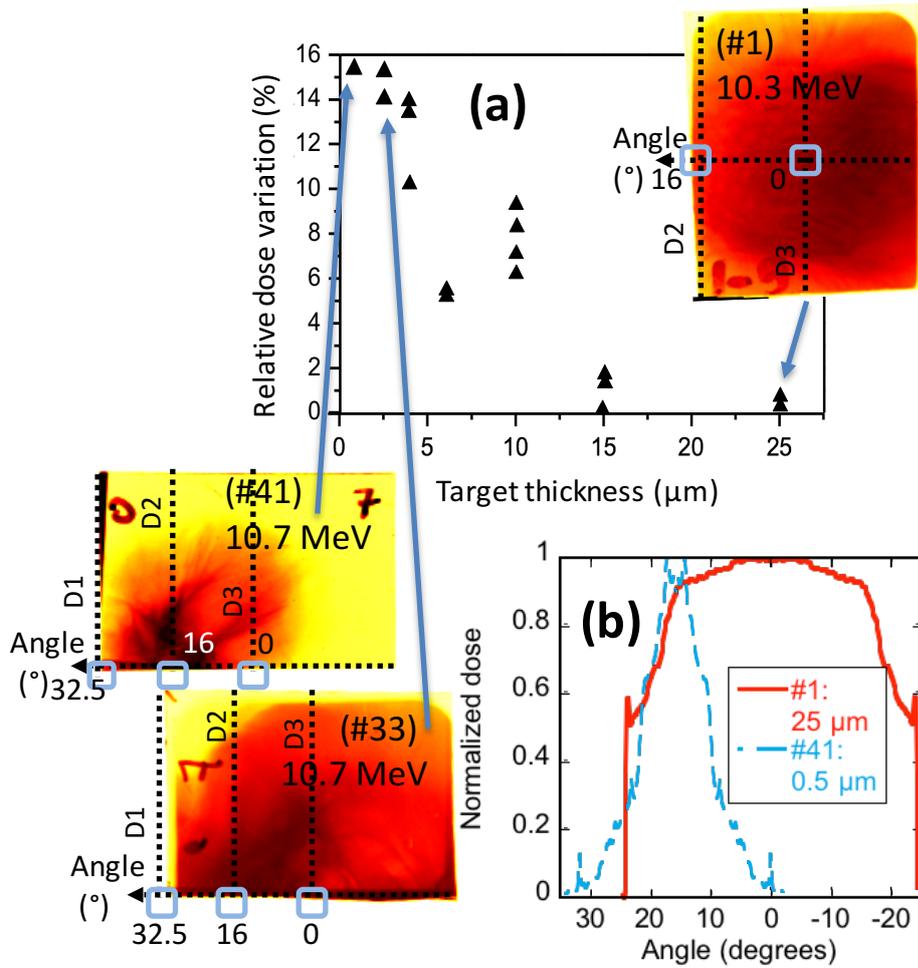

*Figure 4:* *(a) Relative dose variation as measured across the proton beam recorded using RCF films (as done in Ref. [62]), for various target thickness. The targets are of metal for thicknesses ≥ 4 µm, and Mylar for thicknesses ≤ 4 µm; they are irradiated by the short-pulse and its ASE. The more homogeneous the proton beam dose distribution in a film is, the lower the relative dose variation. Three sample RCF images are also shown. For the partially decompressed targets, i.e. that still have a steep rear density profile, and for which TNSA is predominant, are shown a film from a shot (#1) using a solid 25 µm gold target (right) and one from a shot (#33) using a Mylar 2.5 µm target (left, bottom; that RCF was positioned only on the top half of the proton beam). One sample RCF image for the case of a 500 nm Mylar foil shot (#41) where LDCSA is at play is also shown (left, top). Each RCF has its contrast enhanced to show details (the color map goes from dark [highest dose] to yellow [background]). All RCF were positioned such that they were horizontally centered along the target normal (the angle 0°, see Fig.1.a). The horizontal dashed lines on the RCFs indicate the equatorial plane of the experiment (note that for shot #1, the RCF was placed fully in front of the proton beam whereas for shots #33 and #41, the RCFs were intercepting only the top-half of the proton beams in order to leave free the spectrometers line-of-sight). The 0° angle corresponds, as indicated in Fig.1, to the target normal direction. The vertical dashed lines on the RCFs materialize the angular locations of the spectrometers D1, D2 and D3, whose line-of-sights are further marked by empty blue squares. One clearly sees that the proton beam, which is centered at 0° when using the thick targets, progressively shifts*

*angularly toward the left as the target thickness is decreased and the target becomes of lower density under the action of the ASE (see Fig.1.b). (b) Azimuthally-averaged angular lineout of the normalized dose of the RCFs of shots #1 and #41, as indicated. The thick target displays a broad angular profile, as expected for a TNSA-generated beam [10]. On the contrary, the thin target, for which LDCSA is at play, obviously displays a much narrower angular profile which is clearly off-centered around 16°.*

Finally, we have investigated and compared the spatial profiles of the proton beam generated by the exploding target and solid Au foils. The assessment of the beam homogeneity has been done by calculating the mean dose of the beam imprint in the RCF and then computing the variation of the dose with respect to that mean dose, similarly as discussed in Ref. [62]. Figure 4 shows the different values of relative dose variation for various thicknesses of targets (Au and Mylar), either in the TNSA or LDCSA regimes. One can see that the thinner the targets, the less smooth the proton profile is, but also that the proton beam quality still compares to what can be obtained in the TNSA regime (see also similar measurements in the TNSA regime in Ref. [62]).

The next steps are to use over-critical density gas jets [63,64] which are readily available and will offer potential for high-repetition rate use, eliminating the need for solid target replacement. With existing near-infrared (0.8-1 μm wavelength) lasers developing intensities in the range of $10^{21}$ W/cm$^2$, and with newer facilities that will offer [65] even higher on-target intensities, simulations have shown that the 100 MeV barrier could be reached [38,66]. This shows that LDCSA holds promise for accelerating ions to high energies and in collimated manner, while at the same time eliminating the stringent constraints imposed both on targets and laser conditions compared to other novel acceleration schemes (e.g. RPA).

**Author Contribution**
J.F. and E.d.H. conceived the project. P.A., J.F., S.N.C., J.B., M.G., L.R., H.P. performed the experiments, with support from J.C.K., M.St. and O.W. P.A., J.F., M.G. M.B., F.C., M.Sc. analysed the experimental data. E.B. and L.O.S. performed the PIC simulations. S.N.C., J.L.F. P.N., V.T. performed the hydrodynamics simulations. J.F., S.N.C., E.B., E.d.H. and P.A. wrote the bulk of the manuscript, to which all authors contributed.

**Additional Information**
The authors declare that they have no competing financial interests.

**Acknowledgements**

The authors thank the staff of the Titan Laser and the Jupiter Laser facility for their support during the experimental preparation and execution. We thank R. C. Cauble for discussions, P. Combis of CEA for the use of Esther, and G. Revet for discussion on the hydrodynamic simulations. This work was partly done within the LABEX Plas@Par project and supported by Grant No. 11-IDEX-0004-02 from Agence Nationale de la Recherche. It has received funding from the European Union's Horizon 2020 research and innovation programme under grant agreement no 654148 Laserlab-Europe, and was supported in part by the Ministry of Education and Science of the Russian Federation under Contract No. 14.Z50.31.0007. The use of the Jupiter Laser Facility was supported by the U.S. Department of Energy by Lawrence Livermore National Laboratory under Contract DE-AC52-07NA27344. This work is supported by FRQNT (nouveaux chercheurs, Grant No. 174726, Team Grant 2016-PR-189974), NSERC Discovery Grant (Grant No. 435416), ComputeCanada (Job: pve-323-ac, P. Antici). E.B. and L.O.S. were supported by the European Research Council (ERC-2010-AdG grant no. 267841). Simulations were performed on the Accelerates cluster at IST, on the ComputeCanada cluster, on the supercomputers Juqueen (Julich supercomputing center, Germany) and Fermi (Cineca, Italy), through Prace allocations and at the Accelerates Cluster (Lisbon, Portugal). O.Willi acknowledges the DFG Programmes GRK 1203 and SFB/TR18.